\documentclass{elsart}
\usepackage{natbib}
\usepackage{epsfig}
\usepackage{unites2e}
\begin{document}
\runauthor{Colin}
\begin{frontmatter}
\title{Optimization of large homogeneous Air Cherenkov Arrays and application 
to the design of a 1TeV-100TeV $\gamma$-ray observatory.}

\author[UU,MP]{P. Colin}
\author[UU]{and S. LeBohec}
\address[UU]{Department of Physics, University of Utah  Salt-Lake-City, UT, 84112-0830, USA}
\address[MP]{now at the Max-Planck-Institut f\"ur Physik, 80805 M\"unchen, Germany}

\begin{abstract} 
At the time large air Cherenkov arrays are being discussed for future $\gamma$-ray
observatories, we review the relationship between the targeted capabilities
and the main design parameters taking into account construction
costs. As an example application, we describe a telescope array optimized for
observations between 1\,TeV and a few 100\,TeV and use detailed simulations 
to estimate its performances in comparison to science objectives. 
 
\vspace{1pc} 
\end{abstract} 
\end{frontmatter}

\section{Introduction}
\begin{figure}
  \includegraphics[width=7cm]{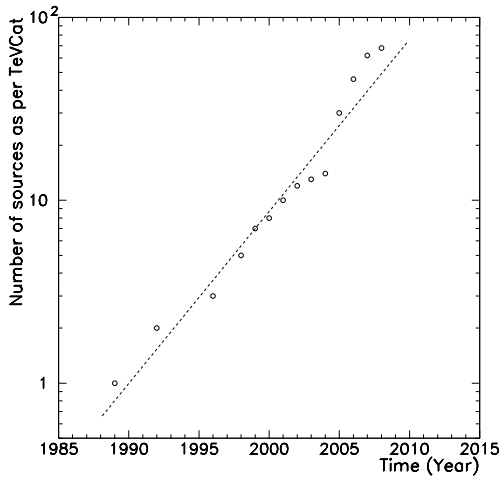}
  \caption{The number of established VHE $\gamma$-ray sources as a function of time
 according to TevCat \cite{tevcat08}. The straight line is indicative of a 3.2 years doubling time.
}
 \label{kifuneplot}
\end{figure}

Gamma-ray astronomy at more than $\rm 100\,GeV$ has entered a
phase of explosive development as illustrated in Figure \ref{kifuneplot} which
shows a source count doubling time just a little larger than 3 years. Possibly more
interesting than the net number of sources is the diversity in the nature of
objects found to produce Very High Energy (VHE) radiation. The VHE $\gamma$-ray
sky now counts various flavors of Active Galactic Nuclei, X-ray binaries,
shell type supernova remnants, pulsar wind nebulae, a pulsar, galactic diffuse
emission, star forming regions and a plethora of sources that still have to be
identified \cite{aharonian2008}. For the most part, these recent successes
resulted from the deployment of Imaging Air Cherenkov Telescope (IACT) 
arrays which provide the 
highest angular resolution and instantaneous sensitivity at these energies.
This motivates the world-wide ongoing efforts toward the 
construction of new large IACT arrays designed to achieve one order of 
magnitude improvement in sensitivity while broadening energy 
range coverage. IACT arrays now in operation count from 2 to 4 large 
telescopes and the targeted sensitivity improvement implies much larger 
facilities which have to be carefully optimized to obtain a maximal science 
capability out of always limited budgets.

In this paper, we discuss the optimization of a uniform IACT array and,
as an example, we present an array to concentrate on the 1\,TeV-100\,TeV
energy range. In section 2 we present considerations for the design and 
optimization of an array based on the properties of the atmospheric Cherenkov
radiation emitted by VHE $\gamma$-ray air showers. This can be used to draw 
prescriptions for large IACT-array main design parameters given the energy range
and collection area to be achieved while minimizing costs. In section 3 we 
discuss the benefits from improving sensitivity at energies greater than 
1\,TeV and present the performance requirements to effectively extend 
$\gamma$-ray astronomy to $\sim$100\,TeV. This motivates the design of the 
example IACT array presented and characterized in section 4 with detailed 
simulations.  

\section {Considerations for the design of IACT arrays}

\subsection{Design and performance key parameters}
The design of an IACT array should be driven by the science objectives and the
corresponding  performances to be achieved. The primary capabilities 
of a telescope array are the $\gamma$-ray effective collection area, the 
cosmic-ray (CR) background rejection performance and the
angular resolution over the covered energy range and effective field of view.
In great part and in various ways, this sets the sensitivity of the 
observatory.
These performance characteristics are related to the array-design 
parameters in a way that depends on the properties of the atmospheric
$\gamma$-ray-shower Cherenkov radiation as investigated here.
In order to limit the number of design parameters, we chose to restrict our 
study to arrays made of a single type of telescopes, uniformly distributed on 
a periodic lattice, and using identical homogeneous cameras. The study remains
applicable to observatories made of several homogeneous sub-arrays specialized
in different energy bands for example. The primary design parameters defining 
such homogeneous arrays are: the number of telescopes, $N$, the
elementary-cell shape, the inter-telescope distance, $\Delta T$, the 
light-collector
diameter, $d$, the field of view, $\psi$, and the pixel size, $\omega$.
The overall photodetection efficiency (including the mirror reflectivity, light 
cone efficiency, photodetector quantum efficiency, etc.) could be also added 
in this list but it is generally technology limited and cannot be considered 
as a real free parameter. Moreover, an increase in photodetection efficiency 
would result in the same improvement as a corresponding increase in the 
telescope diameter, thus, $d$ can be regarded as an effective telescope 
diameter.

All performances of an array can not realistically be addressed 
with precision without detailed simulations such as presented in section 4 for 
an example array. Exploiting geometrical considerations and the properties of 
the atmospheric Cherenkov light from VHE $\gamma$-ray showers, it is however
possible to identify relationships between the energy range 
($E_{Min} -E_{Max}$), the effective collection area, $A_\gamma$, and the 
design parameters of a uniform telescope array. These relations can serve as
guidelines in the design of IACT arrays. There is nevertheless not a one to one
relation between performance and design parameters. An external constraint must
be used to make a choice and that constraint is most likely of financial 
nature. For that reason, after exploring the connections between design and 
performances, we will consider cost issues to obtain a prescription for the 
design of large IACT arrays. In this process, we assume a minimal number
of telescope participating in an event is required for a good reconstruction,
but we ignore the possible dependences of $\gamma$-ray/CR discrimination and
angular resolution on other design parameters that can realistically be addressed
only with detailed simulation of the instruments.
 
\subsection{Effective collection area and Array foot-print}
Stereoscopic observation with IACT has proven to be the best technique to 
reconstruct the direction and energy of primary $\gamma$ rays. This technique 
relies on multiple views of a single shower from several positions on the 
ground. Without considering any specific properties of the atmospheric 
Cherenkov light, it is clear that stereoscopic reconstruction requires the 
shower axis to be at a distance from the telescopes that is not too large 
compared to the inter-telescope distance. Hence the effective area $A_\gamma$
of a large IACT array is larger but close to the geometrical area covered by 
the array foot-print.

\begin{figure}
  \includegraphics[width=13cm]{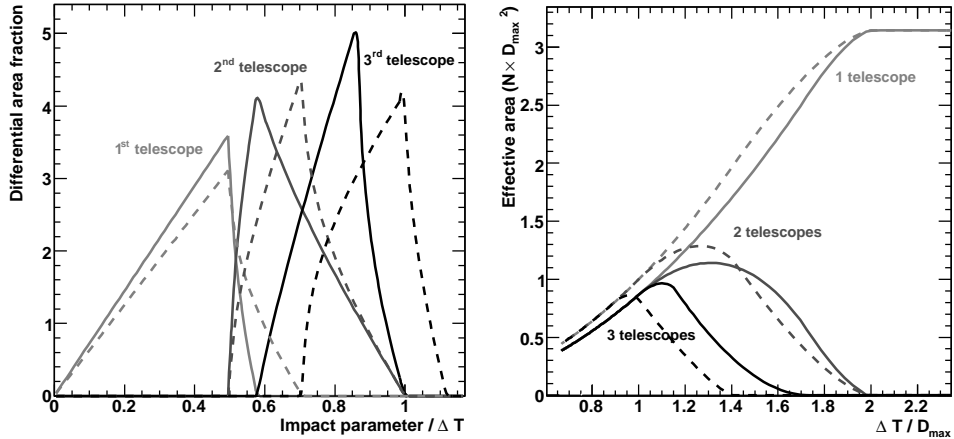}
  \caption{ The left panel shows the distribution of the distance from points 
inside a telescope array to the first, second and third closest telescopes in
the cases of hexagonal (solid line) and square (dashed line) arrays for zenith
observations (inter-telescope distance is used as distance unit).
 The right panel shows the area at less than a distance $D_{Max}$ to one, two
 and three telescopes covered by a $N$-telescope array with inter-telescope 
distance $\Delta T$ and hexagonal (solid line)  or square (dashed line) lattice 
as a function of the ratio $\Delta T$/$D_{Max}$.}
  \label{cellshape}
\end{figure}

In an homogeneous array, the choice of the elementary cell shape affects 
how large an area a $N$-telescope array covers. Using the inter-telescope
distance $\Delta T$ as a unit, Figure~\ref{cellshape} (left) shows the distribution 
of the distance from the shower axis to the closest three telescopes inside 
hexagonal and square arrays respectively.
The effective collection area at a given energy depends on the maximal 
distance, $D_{Max}$, guarantying a single-telescope detection and on the 
minimal number of telescope required to participate in an event.
For a given $D_{Max}$ (We will later investigate how this distance arises),
the effective area scales as $\Delta T^2$ as long as $\Delta T$ is much smaller than $D_{Max}$.
For larger $\Delta T$, the full geometrical area may not be effective.
If a single-telescope detection is required, the effective area would rise up to
$N \cdot \pi \cdot D_{Max}^2$ at $\Delta T = 2 D_{Max}$ (no more overlap) and
then remains constant.
When a multi-telescope detection is required, $A_\gamma$ reaches a maximum and
then decreases to 0 when $\Delta T > 2 D_{Max}$.
Figure~\ref{cellshape} (right) shows the effective area as a function of the 
ratio $\Delta T$/$D_{Max}$, for a single, double and triple telescope
detection requirement in hexagonal and square arrays for zenith observations.

Stereoscopic observation requires at least two participating
telescopes. The two telescopes and the shower axis can however be in a
same plane in which case a degeneracy arises in the reconstruction.
Having images recorded from three non-aligned ground positions guarantees a 
good stereoscopic reconstruction is always possible and can be considered as
a better minimal requirement. Having more than three telescopes 
participating in an event is beneficial to both the angular resolution
and CR discrimination. However, increasing the telescope multiplicity 
requirement reduces the effective collection area.
Here, we investigate the optimization of the effective area at the energy threshold,
so we choose a low multiplicity level at this energy.
At higher energies, the telescope multiplicity increases and the associated
advantages become available.

We choose $\Delta T$ to maximize the effective area at the targeted energy
threshold. For large $N$, the array area scales as $N\cdot \Delta T^2$ for a square 
lattice and as $N\cdot \sqrt{3}/2 \cdot \Delta T^2$ for a hexagonal lattice.
So, for a given $N$ and $\Delta T$, a square array would cover a larger area than a
hexagonal array but the hexagonal array preserves telescope multiplicity for larger
inter-telescope distances. For a triple-multiplicity, the optimum configuration is
a hexagonal array with $\Delta T$\,=\,1.1\,$D_{Max}$ (right side of Figure~\ref{cellshape})
which provides a $12\%$ larger effective area than the maximal square-array area with the same number
of telescopes (and so, for a same price). With a 2-telescope requirement, the optimum configuration
is a square array with $\Delta T$\,=1.265\,$D_{Max}$.

\begin{figure}
  \includegraphics[width=6.5cm]{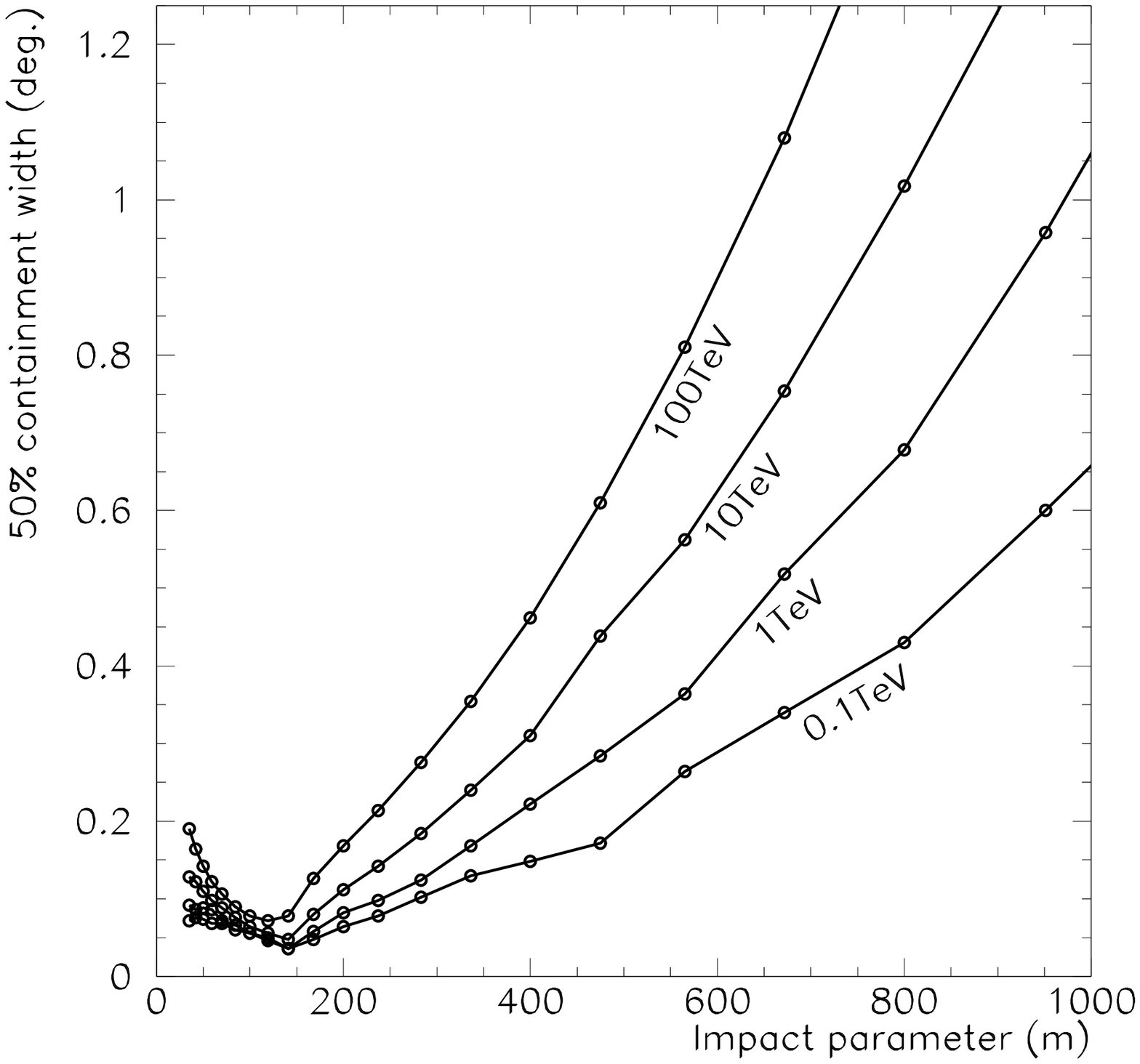}
 \includegraphics[width=6.5cm]{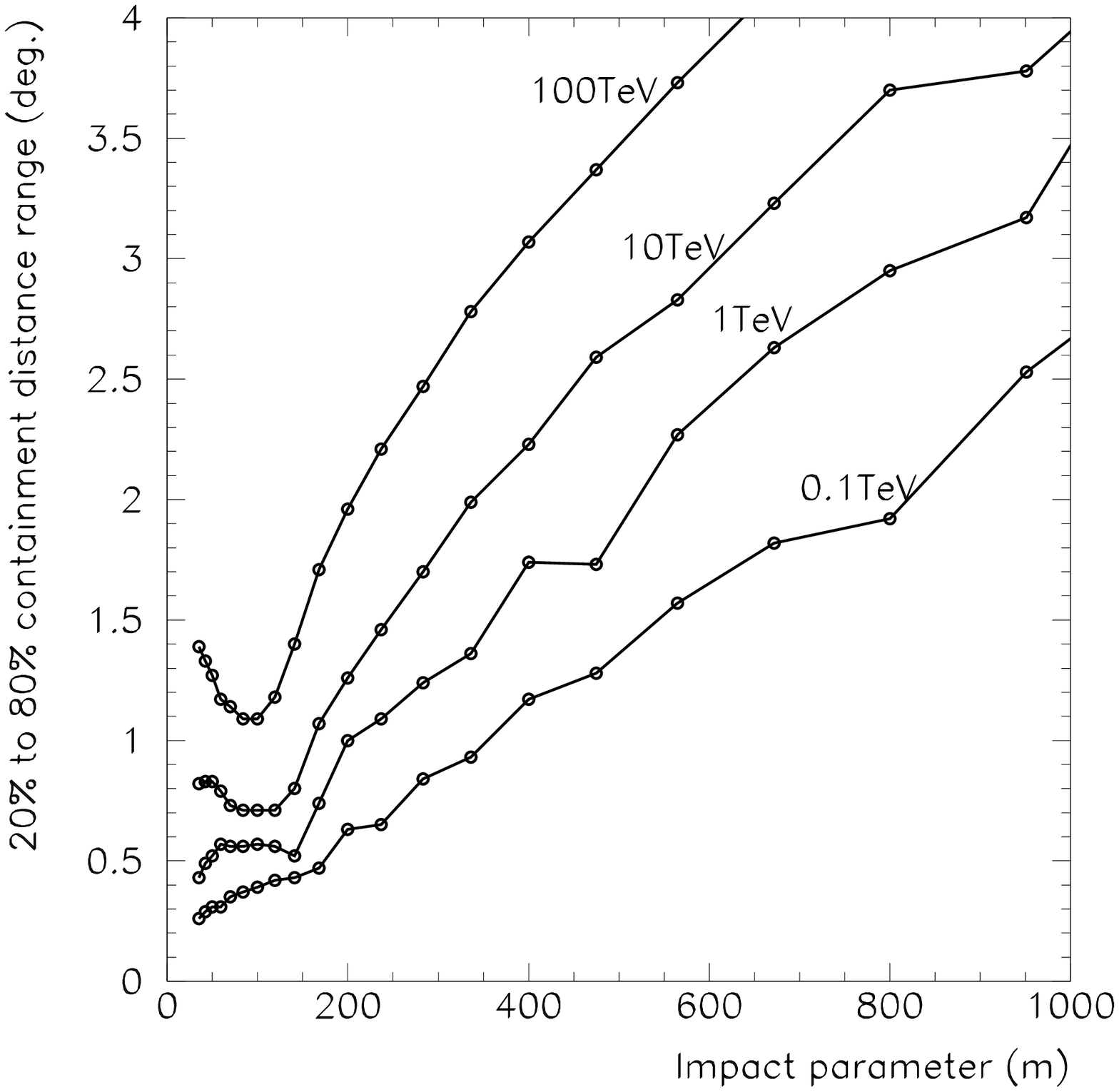}
  \caption{ The 50\% containment $Width$ (left) and the 60\%
  containment $Length$ (right) of average air Cherenkov $\gamma$-ray image of 
  few showers are shown as a function of the impact parameter to the 
  telescope for
  observation at $30^\circ$ from zenith from a site 1270\,m above sea level.}
  \label{width}
\end{figure}

\subsection{Pixel size }\label{pixsiz}

The characteristics of IACT camera (field of view, pixel size), must match
the properties of VHE-$\gamma$-ray-shower Cherenkov images in the range of 
energy ($E_{Min} - E_{Max}$) and range of impact parameters to telescopes
within a cell (0 - $D_{Max}$) for the given IACT-array design (position and 
size of the telescopes).
The properties of Cherenkov images are obtained from 
Monte Carlo simulations with KASCADE \cite{kascade} of $\gamma$-ray showers for energies 
from 100\,GeV to 100\,TeV, at 1270\,m above sea level, for observations 
$30^\circ$ from zenith to the north and to the south. Image properties obtained
to the North and to the South can be slightly different because of the effects
of the geomagnetic field (southern Arizona in our simulation). Here, we use an 
average of the North and South profiles. $\gamma$-ray-shower images have a 
regular shape, elongated along the direction of development.
They are often characterized by their $Length$ and $Width$, both depending on
the primary energy and impact parameter to the telescope.  Here, the $Width$ is
defined as the width of a stripe containing 50\% of the Cherenkov light along 
the shower axis. Similarly, the $Length$ is defined as the length of a 
longitudinal section of the image containing 60\% of the Cherenkov light
(Specifically, the $Length$ is calculated as the difference between the 20\% 
and 80\% Cherenkov-light-containment distances from the $\gamma$-ray-source
position in the image).

Figure~\ref{width} illustrates the variation of $Length$ and $Width$ with the 
primary energy and impact parameter. It should be noted that the $Length$ 
and $Width$ as defined here are intrinsic to the shower images while the 
corresponding image parameters in real data analysis can be strongly affected 
by the telescope performance and the level of night sky background.
Hence the actual values are essentially indicative only.
The pixel diameter, $\omega$, must be small enough to resolve the global 
image shape. More specifically it must allow to resolve the major axis direction and
the longitudinal profile of the image.
The pixel size must also be chosen taking into account signal 
to noise ratios at the trigger and image analysis levels. The night sky 
background noise level in a single pixel scales as $\omega$. For a given 
shower image, as long as $\omega$
remains smaller than the $Width$, the single pixel signal scales as $\omega^2$.
For $Width<\omega<Length$, it scales as $\omega$ and for $\omega>Length$, 
the signal does not depend much on the pixel size. This is overly simplified. 
It however remains, that for $Width<\omega<Length$, the signal to noise ratio 
does not depend much on the pixel size and this ratio is degraded for both 
larger and smaller sizes. In order to maximize image analysis capabilities 
without degrading the detection signal to noise ratio, one will likely favor 
pixels whose diameter corresponds to the image $Width$. The pixel size should
match the $Width$ of the most compact shower images which are those at the
targeted energy threshold $E_{Min}$ in the closest telescope
(the distribution of which peaks at 0.5 $\Delta T$ in the left side of 
Figure~\ref{cellshape}). This provides a first good idea about the range of 
pixel sizes that should be considered in detailed simulations.

\begin{figure}
  \includegraphics[width=7cm]{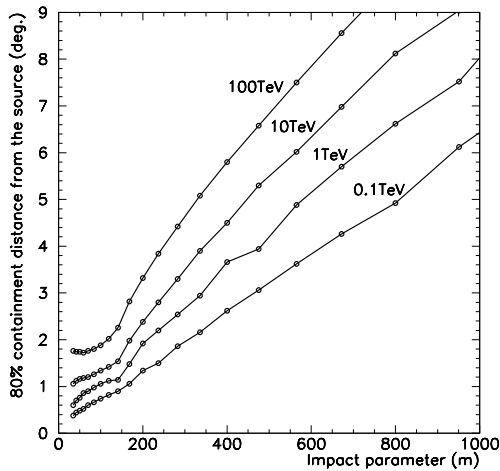}
  \caption{The 80\%-Cherenkov-$\gamma$-ray-shower-image-light-containment distance
  from the source is shown as a function of
  the impact parameter for a range of energies at $30^o$ from zenith from a 
  site at 1270\,m above sea level.}
  \label{anglext}
\end{figure}

\subsection{Camera field of view}
The camera field of view, $\psi$, must be large enough to contain shower
images of interest.
Figure~\ref{anglext} shows the 80\%-Cherenkov-light-containment angle from
the $\gamma$-ray-source position as a function of the impact parameter and for
different energies. The greater the energy, the faster the containment angle
increases with the impact parameter. Thus, in order to guaranty a
$p$-telescope multiplicity requirement, $\psi$ must be chosen so images of 
showers at the maximal targeted energy $E_{Max}$ are still contained
for  impact parameters to the $p^{th}$ closest telescope.
For a triple multiplicity, the third closest telescope in a hexagonal array lies at $\sim$0.87\,$\Delta T$
(left side of Figure~\ref{cellshape}). On top of that minimal field of view, 
the choice of $\psi$ must also give some provision to allow for the effective
observation of extended or poorly localized sources. Even in the case of 
point-like sources, a large enough effective $\gamma$-ray field of view, 
$\psi_\gamma$, is advantageous for a good background level characterization.
Here, $\psi_\gamma$ is the field within which $\gamma$-ray images
are not truncated at the edge of the optical field of view.
In existing IACT array, $\psi_\gamma$ was chosen between 1$^\circ$ and 
2$^\circ$. Future observatories could benefit from larger field of views
in order to improve their sky-survey capability for example.

Another advantage of cameras with larger field of view is that they allow 
showers falling further away outside the array geometrical area to still be 
detected. Then, the effective area increases with energy, from close 
to the array foot-print at the energy threshold, up to a maximal area at the 
energy for which field truncation effects become too important.
It is therefore possible to obtain a larger effective area at the highest 
energies by placing a few larger field of view telescopes on the outer edge 
of the array. Here, however, we will not study this option as we  restrict 
ourselves to arrays in which all the telescopes are identical.

\subsection{Energy threshold}

\begin{figure}
  \includegraphics[width=7cm]{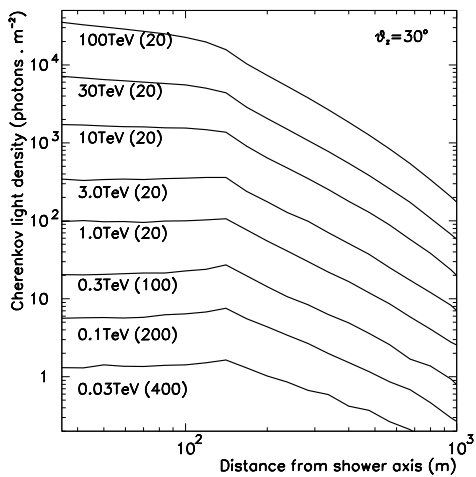}
  \caption{The Cherenkov-light density at ground level from $\gamma$-ray showers
  are shown for a range of energies as a function of the impact parameter for
  observation at $30^\circ$ from zenith from a site 1270\,m above sea level. The atmospheric extinction is taken into account and the considered wavelength range is 250-700\,nm.}
  \label{cherdensit}
\end{figure}

The capability of an IACT to detect an air shower depends primarily on the
effective light-collector area, on the Cherenkov-light density at
ground level and, to some extent, on the angular extension ($Length$ and $Width$)
of the shower image. Figure \ref{cherdensit} shows the density of Cherenkov light 
projected on the ground by $\gamma$-ray showers at $30^\circ$ from zenith 
as a function of the 
distance to the shower axis for several primary energies. The Cherenkov-light 
density is relatively uniform in a 150\,m-radius plateau and then decreases 
rapidly. 
The break at 150\,m results from the geometric properties of Cherenkov light 
combined with the atmospheric density profile. The radius of the Cherenkov light pool plateau slowly decreases with the altitude \cite{lebohecholder2003} and strictly speaking these curves only apply directly for the altitude for which simulations were done (1270\,m).
The tail extending beyond the break is due to shower particles undergoing various amounts of multiple 
scattering. In principle, this tail can be used for detecting showers with 
large impact parameters. As the Cherenkov-light density in the tail decreases 
with increasing of impact parameter, the energy threshold of an array
increases with increasing inter-telescope distance. This motivated 
the idea that, in order to extend coverage to the highest energies,
one should or could consider increasing the inter-telescope distance in the 
periphery of the array (as in circulating possible designs of CTA).
However, it appears the energy threshold of a sparse array can be achieved with 
a higher density of smaller telescopes which might turn out to be less 
expensive. In order to investigate this, we must identify the different ways
in which the energy threshold arises.

\begin{figure}
  \includegraphics[width=7cm]{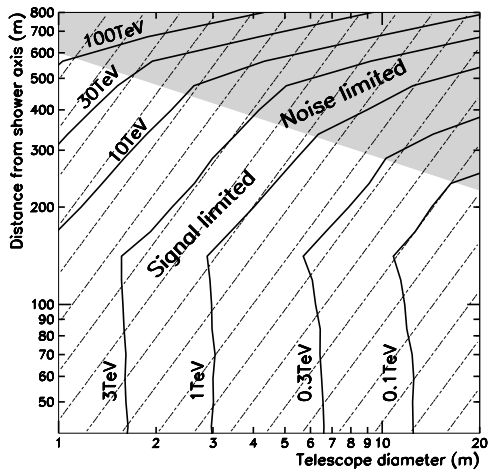}
  \caption{The solid lines are the iso-energy-threshold curves 
calculated with a simple model as a function of the telescope diameter and 
distance to the shower axis. Signal and a noise limited threshold regions are 
highlighted. The dashed lines are iso-cost curves for arrays assuming the price is
proportional to $d^{2.7} \over \Delta T^2$.
Iso-cost curves are separated by a factor of two.
}
  \label{isothreshold}
\end{figure}

The energy threshold can be signal or noise limited. It can be signal limited 
because when the number of photo-electrons is too small, image
characterization associated with the shower-reconstruction possibilities deteriorates.
It can be noise limited because the signal in each pixel has to stand above 
the surrounding noise for images to be identified and characterized. For a 
given night-sky-background (NSB) light intensity, the curves in Figure 
\ref{cherdensit} can be used to construct single telescope iso-threshold 
curves as a function of telescope diameter and the impact parameter.
Whichever signal limited and noise limited thresholds is, the highest should be 
considered as the threshold. A set of iso-threshold curves is presented on 
Figure~\ref{isothreshold} for a photo-detection efficiency of 15\% (250\,nm-700\,nm bi-alkali
photocathode quantum efficiency average for the Cherenkov spectrum including atmospheric attenuation effect),
typical of IACT equipped with photomultiplier tubes.
Both the signal and noise threshold limitations have to be set somewhat arbitrarily. Here, the signal 
limited threshold corresponds to a minimum 50 photo-electrons detected by the 
telescope. The noise limited threshold is set by requiring pixels in the image 
core to have an average signal that exceeds NSB fluctuations by a factor of
5. We consider the image core as the part of the image inside a central
$Width \times Length$ rectangle. The individual pixel signal is then estimated
using the argument outlined in the discussion on pixel size (section 
\ref{pixsiz}). The NSB luminosity is taken to be 
200\,photo-electrons $\cdot$m$^{-2}$\,$\cdot$ns$^{-1}$\,$\cdot$sr$^{-1}$.
For the noise contamination, we assumed a 20\,ns integration time.
The choice of the camera design can affect the iso-threshold curves.
Here, we consider a fixed pixel size $\omega$=0.2$^\circ$, which is within the
optimal signal to noise ratio range ($Width<\omega<Length$) identified in
subsection \ref{pixsiz}.

On each iso-threshold curve two breaks appear. The lowest one results from the
break in the distribution of Cherenkov-light ground density at 150\,m.
The highest one corresponds to the transition from the signal to noise limited
regime. The signal and noise limited iso-threshold curves shift with respect
to one another depending on the specific assumptions made in the definition
of these thresholds. However the overall shape of the diagram remains the
same with a noise limited threshold in the upper right region of the graph.
It would, in principle, be desirable to place oneself close to the boundary
between these two regions, but, 
as we will see, this might not be optimal once budget considerations are taken
into account. Using Figure \ref{isothreshold} and assuming a $p$-telescope
multiplicity requirement, one can use the typical distance to the $p^{th}$
closest telescope (Left side of Figure~\ref{cellshape}) in order to get an 
idea of the energy threshold of a large array with a given telescope diameter $d$
and inter-telescope spacing $\Delta T$.

\subsection{Costs considerations}
To a first approximation, ignoring the cost associated with the infrastructure,
the overall cost of an array scales linearly with the number of telescopes and
hence, for a chosen array area, with the inverse inter-telescope distance 
squared ($ N\propto$ $1 \over \Delta T^2$).
The individual telescope unit price is made up of the light collector
(on its mount and pedestal) and the camera. The light-collector price is often
considered to scale as a power law of the diameter with an
index $\alpha$ between 2.5 and 3. Up to now, the prices of imaging air
Cherenkov cameras have been of the same order as that of the light-collectors
they were for, so the single telescope price
roughly follows the same power law as the light collector alone.
In this approximation, we have an array price which scales as $d^\alpha \over \Delta T^2$.
This simple scaling might not apply uniformly over the entire telescope diameter
range covered in figure~\ref{isothreshold}. The materials and technology required
for a 20\,m telescope are likely to be different from that required for a few-meter
telescope. However, this simple model is sufficient to obtain locally
meaningfull iso-price curves. On figure \ref{isothreshold}, we have indicated such
iso-price curves for doubling prices with $\alpha=2.7$. Prices increase perpendicularly
to the iso-price lines toward the lower right (greater numbers of larger telescopes).
Following one iso-threshold curve up from the bottom of the graph, one sees the price
decreases rapidly until one reaches the first break at 150\,m. Then the price starts
to slowly increase or, for the lowest energies (see 0.3\,TeV and 0.1\,TeV), it remains
constant until one enters the noise limited threshold region where the price increases
rapidly. This clearly indicates the inter-telescope distance of an array should be chosen
in such a way that the largest impact parameter necessary to satisfy the telescope
multiplicity requirement is between the two breaks on the chosen iso-threshold curve.
Once this largest impact parameter is chosen, the required telescope diameter can be read
on the graph. For low energy arrays (100\,GeV), the second break might be
preferable as it does not change the price but reduces the number of telescopes to
operate and maintain. For higher energy arrays, the first break at 150\,m is more
attractive as a price difference appears. 

This was without considering the costs associated with the field of views.
IACT arrays in operation have field of views from $3.5^\circ$ to $5^\circ$
which are sufficient for the observation of 10\,TeV showers up to $\sim$200\,m
distances. For large arrays, if larger inter-telescope distances or higher
maximal energies are considered, the necessary field of view increases
(see Figure~\ref{anglext}).
This results in a price increase as the number of channels per telescope camera increases and the
telescope design must ensure an acceptable optical-point-spread function over the entire field of view.
As a consequence, the iso-price curves of figure~\ref{isothreshold} must bend to the left when
considering increasing impact parameters and the correspondingly increasing
field of view requirements.
This makes the second break in the iso-threshold curve even less attractive.

Thus, it seems the optimal array is such that, at the targeted energy
threshold, the maximal impact parameter $D_{Max}$\,=\,150\,m. 
In section 2.2, we showed that for triple-multiplicity requirement,
a hexagonal lattice array with an inter-telescope distance 
$\Delta T$\,=\,1.1\,$D_{Max}$ provides the largest effective area (for a given 
number of telescopes) for Zenith observation.
In practice, observations are made at some angle from Zenith and one dimension 
of the array is reduced by projection effect. For typical observations 
at 30$^\circ$, this projection effect amounts to $\sim$14$\%$. The optimum 
$\Delta T$ should consequently be increased by an intermediate factor 
($\sim$7$\%$). A hexagonal array with 
$\Delta T$\,=\,1.1$\times$1.07$\times$150\,m $\simeq$\,175\,m
appears as the most economically attractive design. In the case of a 
two-telescope multiplicity, following the same argument, the optimal design 
is square lattice with $\Delta T$\,$\simeq$\,200\,m.

\section{Extension of $\gamma$-ray astronomy to higher energies}

\subsection{Motivations}
The prime motivation for the development of $\gamma$-ray astronomy was that 
$\gamma$ rays should trace high energy hadron-CR interactions, 
especially near their still to be identified acceleration sites.
As the VHE $\gamma$-ray source catalog started to grow, the role played by 
inverse Compton interactions of high energy electrons became preponderant.
As of today, we still do not have one source in which the VHE $\gamma$-ray 
emission can be unambiguously attributed to hadron interactions. 
In particular, this is the case with supernova remnants (SNRs) which are 
still considered among the most likely CR accelerators. The 
hadron picture suffers from the $\gamma$-ray emission morphology not matching 
the interaction target material densities. On the other hand, the inverse 
Compton picture suffers from not accurately predicting the observed spectra 
(RXJ1713.7-3946 \cite{RXJ1713b}) or from implying magnetic fields of
magnitude too weak to account for tight confinement of the emission region
(Vela Junior\cite{VelaJr}). 

The soon to come measurement of the $\pi_0$ bump in SNRs with the Fermi-LAT 
$\gamma$-ray space telescope \cite{fermi_lat} should clarify whether the high 
energy $\gamma$ rays are from hadron origin. However, even if the TeV 
$\gamma$-ray emission from RXJ1713.7-3946, for example, is confirmed to result 
from freshly accelerated hadrons, the softening of the $\gamma$-ray spectrum 
above few TeV \cite{RXJ1713b} makes it a non typical CR source.
Spectrum of hadrons accelerated by RXJ1713.7-3946 should show a similar softening at an
energy which is a factor $\geq$20 short of the knee energy. The actual CR spectrum indicates
that major contributors to galactic CR should display pure power law $\gamma$-ray spectra
up to several hundreds of TeV. In order to identify CR accelerators operating up to the knee 
energies, the domain covered by the $\gamma$-ray astronomy has to be extended up 
to a few hundred TeV.

At such high energies, the inverse Compton contribution 
should be strongly suppressed as even the scattering on CMB photons occurs in the relativistic 
regime making hadron processes easier to identify. The relation between the
source morphology and the spectrum at these energies can also help the
process identification. High energy hadrons need larger regions to be accelerated.
So, $\gamma$-ray spectra of hadron accelerators should be harder at larger 
scales. In leptonic models, it is the opposite. Electrons lose more energy
by synchrotron cooling at higher energy and cannot travel as easily on large 
scales. The $\gamma$-ray spectrum should be softer at larger scales.
It should also be mentioned that, in some objects such as pulsar wind nebulae,
the shorter lifetime of electrons at the highest energies may allow also 
to detect injection time variabilities which are smoothed out at lower 
energies.

At $\rm 100\,TeV$ and above, absorption by the interstellar radiation field
becomes a concern but should remain bearable for galactic sources. A $25\%$
attenuation is expected at $\rm 100\,TeV$ for sources at the galactic center
\cite{Moskalenko06}.
Most of the galactic sources detected up to several tens of TeV by the current 
generation of IACT have relatively hard spectra. Generally, their spectra 
are well described by a power-law with a differential spectral index between 
2 and 2.5, and, in many cases, do not show any cut-off. As all emission 
models predict high energy cut-off, it is particularly interesting to observe 
these objects at higher energies. 

\subsection{Minimal exposure requirements}

\begin{figure}
  \includegraphics[width=13.5cm]{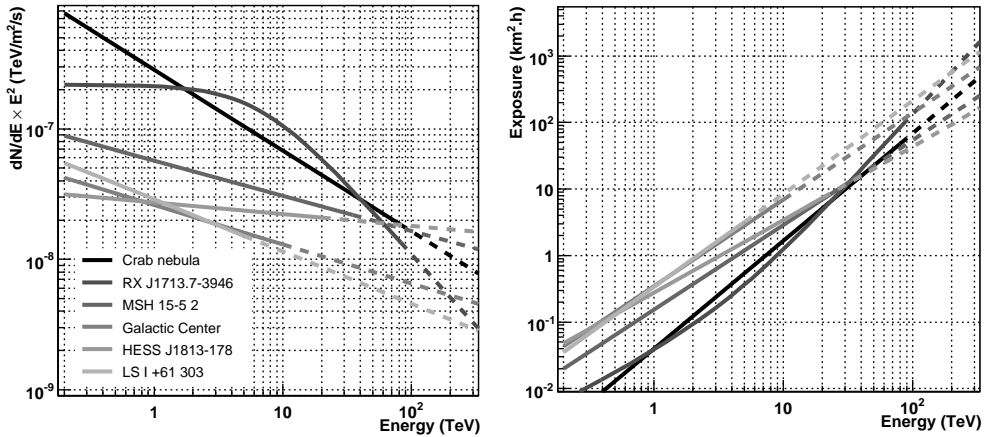}
  \caption{The left-hand side shows the spectral energy distribution of selected galactic sources 
and their power-law extrapolation at higher energy. The right-hand side shows the 
required exposure for detecting 25 $\gamma$~rays from these sources as a 
function of the energy threshold.}
  \label{exposure}
\end{figure}

Observation sensitivity depends primarily on the available exposure, the 
product of the effective collection area $A_\gamma$ with the observation 
time $\tau_{obs}$ dedicated to a given source. The necessary exposure to 
extend $\gamma$-ray-astronomy coverage toward higher energies can be estimated 
from an extrapolation of known $\gamma$-ray-source spectra. One can calculate 
the exposure required to collect enough events above any given targeted energy 
to detect and measure the spectra of these sources. The exposure estimate 
should take into account also the CR background contamination of the signal. 
However, even if at energies higher than few tens of TeV, the CR 
discrimination 
may become less effective, the CR flux decreases rapidly with energy (spectral 
index of -2.7), while most of galactic sources are found to have harder 
spectra. Moreover the $\gamma$-ray angular resolution of IACT improves 
significantly with energy, reducing dramatically the background contamination 
for point-like sources. As we will see in section 4, above 10\,TeV, the 
angular resolution can be better than 2$^{\prime}$. Thus, at least for point sources,  
at the highest energies, the sensitivity may not be limited by the CR 
background but by the number of $\gamma$-ray events itself.

A flux measurement may require more than ten events for statistical
errors to be acceptable. Moreover, half of the $\gamma$ rays are typically
lost in  the background rejection cuts. We chose to set the minimal
exposure by requesting the harvest of at least 25 $\gamma$-ray events.
Figure~\ref{exposure} shows spectra of different types of galactic source 
\cite{RXJ1713b}, \cite{Aharonian06}, \cite{Aharonian04}, \cite{Aharonian04b}, 
\cite{Aharonian05}, \cite{Acciari08} and their high energy extrapolation 
according to the reported spectrum. It also indicates the exposure necessary
to collect 25 events from these sources above any given energy.
This suggests that a 100\,TeV $\gamma$-ray astronomy requires
an exposures of at least $A_\gamma \cdot \tau_{obs} \simeq 100$\,km$^2\cdot$hr.
The proton flux above 100\,TeV is about $2.4 \times 10^{-5}$\,m$^2 \cdot$\,s$^{-1}\cdot$\,sr$^{-1}$
\cite{atic-2}. For an exposure of 100\,km$^2 \cdot$hr, the number of protons 
collected above 100\,TeV in a 2$^{\prime}$-radius sky region ($10^{-6}$\,sr) is less 
than 10. At 100\,TeV, the sensitivity to point-like source is then effectively
more limited by the $\gamma$-ray statistic than by a CR-background contamination.

Other experimental approaches to 100\,TeV astronomy than presented here have
been considered \cite{Colin2007} but they generally provide much inferior 
angular resolution.
The angular resolution achieved by IACT arrays plays an important role in
setting the sensitivity to point-like sources. 
A large portion of galactic sources, however, are already found to be extended.
For those, the angular resolution plays a role that is less important in a discovery mode,
but remains highly valuable as it impacts morphological study capabilities. In fact,
another interesting advantage of extending $\gamma$-ray astronomy to higher energies
actually is the improvement of the angular resolution it provides.

\subsection{Advantages of a 100\,TeV capable array}

Future observatories sensitive to 100\,TeV $\gamma$-ray sources should be
designed with a threshold providing a good overlap with the current IACT
arrays which operate at their best below a few tens of TeV. Among the known VHE 
$\gamma$-ray sources, to the exception of the Crab pulsar and some 
distant AGNs, not one has a spectrum with a cut-off energy below 
1\,TeV. A targeted energy threshold around 1\,TeV seems low enough for a 
good overlap, and allows the detection of a large number of sources.
The next generation $\gamma$-ray astronomy projects will certainly be intensively
used in dedicated studies of the most interesting already known objects.
The higher sensitivity of these observatories will also be used to further
the ongoing exploration of the VHE $\gamma$-ray sky. The increasing number
of $\gamma$-ray sources can be expected to result in a decrease of the observation time 
typically allocated to each one with future projects. Currently,
observing campaigns of one object often count around 50\,hr per year, more 
than 5\% of the usable night sky time. This limits to a few tens the number 
of sources studied each year. The next generation of IACT arrays should 
allow the observation of hundreds of sources every year, reducing to
a few hours the time allocated to each. With large enough telescopes 
to achieve a TeV threshold, the large array area required for the highest energies
would also result in a tremendous sensitivity gain at a few TeV compared to
the present generation of observatories.

\section{Example of a E$>$1\,TeV IACT array}
\subsection{Specific design}
As an example, we design an IACT $\gamma$-ray observatory dedicated to the 
1\,TeV-100\,TeV energy range and study its performance with detailed Monte 
Carlo simulations. Following discussions in section 2, we choose the array to 
be a hexagonal lattice with an inter-telescope distance $\Delta T$\,=\,175\,m.
For a 1\,TeV threshold at impact parameter around 150\,m,
Figure~\ref{isothreshold} indicates the telescope diameter should be about
3\,m. The field of view of the camera must be large enough to well contain 
images of 100\,TeV showers with impact parameters up to 150\,m. 
Figure~\ref{anglext} shows that a minimal $2^\circ$-radius field of view is 
necessary.
Often, for background control purpose, observation with IACT arrays 
are taken in wobble mode with the source position off-centered by typically 
$0.5^\circ$. Thus, we choose a $2.5^\circ$-radius field of view for the 
camera ($\psi=5^\circ$, $\psi_\gamma=1^\circ$).
The $Width$ of 1\,TeV $\gamma$-ray-shower images with impact parameters
below 150\,m is about $0.05^\circ$. Pixels that small would
imply cameras with an unrealistically large number of channels for standard
photomultiplier technology based cameras. We decided to explore the
consequence of relaxing this requirement and completed the Monte Carlo study
for $\omega =0.16^\circ$, $0.22^\circ$, $0.32^\circ$ and $0.42^\circ$ in
order to investigate how performances degrade with pixel size. With a
hexagonal lattice arrangement of pixels, this corresponds to cameras with
respectively 967, 499, 253 and 151 pixels. The optical point spread function
required for such cameras is compatible with a Davis-Cotton optics with an
aperture ratio of $\sim 1$. Thus, we adopt such a design for the light
collectors.

\begin{figure}
  \includegraphics[width=6.5cm]{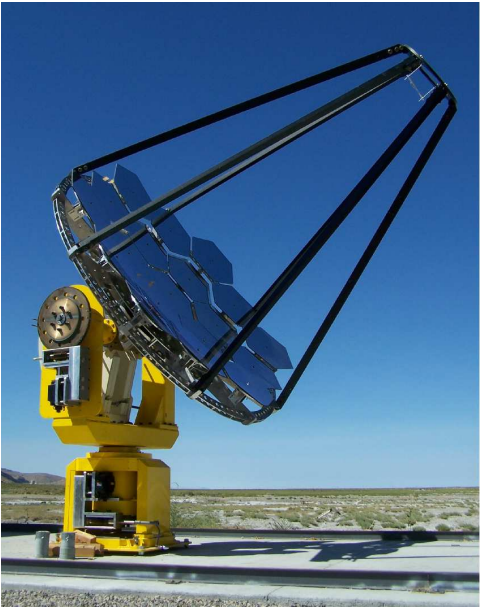}
  \caption{One of two of the 7 TA telescopes, reassembled as part of the Star
  Base Utah observatory for prototyping purpose.}
  \label{starbase}
\end{figure}

For the detailed telescope configuration we used the design of the Utah Seven 
Telescope Array \cite{sevenTA}. These telescopes are 6\,m$^2$ area, f/1 
Davis-Cotton, light collectors made of 19 hexagonal mirror facets
(Figure~\ref{starbase}) with a 3.185\,m focal length. Two of the seven 
telescopes have been recently redeployed at the Star Base Utah Observatory 
\cite{StarBase} near Salt Lake City as a prototyping test facility. According
to Figure \ref{isothreshold}, such an array of small telescopes will 
achieve an energy threshold that is signal limited rather than noise limited. 
This somewhat relaxes the requirements for high speed electronics. As a 
consequence for the electronics simulations, we used the Whipple 10\,m 
electronics model which is very conservative with a 25\,ns Q-ADC integration gate.
At such high energies ($>$1\,TeV), the individual telescope trigger rate is
very manageable and we did not include any telescope coincidence to reduce
accidental rates. However in the analysis, we require a 3-telescope coincidence
to reconstruct the shower.

\subsection{Monte Carlo Simulation and Data Analysis}
\begin{figure}
  \includegraphics[width=7cm]{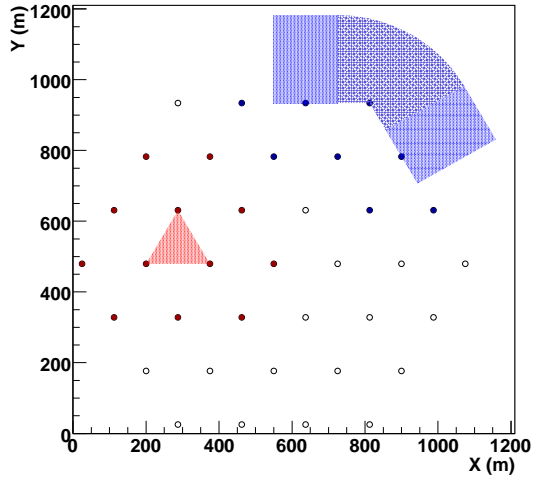}
  \caption{Shower impact regions and simulated sub-arrays.
Red telescopes are considered for the inside showers (Triangle).
Blue telescopes are considered for the outside showers (side and corner regions).}
  \label{array}
\end{figure}

The Monte Carlo simulations are performed using the GrISU(tah) package 
\cite{GrISU}. We are interested in the performance of large arrays.
Because of the relatively small field of view chosen, a telescope 
cannot effectively record images of showers with too large impact 
parameters. With $\psi=5^\circ$,
a 1\,TeV-shower image is well contain in the camera only up to 300\,m (Figure~4) and
this worsens at higher energies.
Most air showers will be detected only by a small fraction of the telescopes. 
Thus, we restrict our simulations to sub-arrays close to the shower-core 
impact point. The inside of a hexagonal array can be subdivided in triangular 
cells. In order to account for showers falling inside the array, we simulated 
showers inside a triangular cell and included the response of only the 12 
closest telescopes around this cell. For the showers falling outside the 
array, we assume that the external shape of the array is a regular
hexagon. The external belt can be subdivided in several regions of two types: 
the ``sides'' (next to telescopes aligned with their neighbors) and the 
``corners'' (next to corner telescopes). Figure~\ref{array} shows the geometry 
of the different region types and the associated simulated sub-arrays. The results
of these simulations can then be combined with different weights to 
characterize any size hexagonal arrays with more telescopes than in the 
simulated sub-array. A simple way to build a hexagonal array is to add
hexagonal rings of telescopes around a central one. With $n$ rings we get 
 $3n(n+1)+1$ telescopes, $6n^2$ triangular cells, $6(n-1)$ side regions 
and 6 corner regions. Here, we present results for arrays with from 19 to 469 telescopes.

With this approach, we are neglecting the fact that showers falling in 
triangular cells on the edge of the array are not surrounded by 12 telescopes 
but by only 9 or 8. However, the missing telescopes are distant from the shower 
and they are not expected to play a central role in the event detection and reconstruction.
On the other hand, at high energy more telescopes than the 12 considered in the simulated
sub-array will trigger. These additional images, even truncated,
may slightly improve the angular resolution and CR rejection.

The Monte Carlo data are analyzed with one of the analysis chains used for the 
VERITAS data. Shower images recorded by each telescope are preprocessed with 
the standard two-threshold-cleaning algorithm. In order to be part of the 
image, a pixel must exceed a high threshold of 4\,photo-electrons or exceed a 
low threshold of 2\,photo-electrons and neighbor a pixel that exceeds the high 
threshold. With such small telescopes, the NSB per pixel and per integration 
gate is typically well below 1\,photo-electron. A cleaned image is considered
usable for event reconstruction when it is made of at least 4 pixels and 
contains at least 25\,photo-electrons. An event is considered if 3 images 
satisfy the usability criteria. We use the same requirements for all pixel 
sizes. These requirements set the analysis-level energy threshold of the
array.
 
The arrival direction of a good quality event is reconstructed by
stereoscopy. For each pair of usable images a direction is reconstructed. 
The final reconstructed direction is obtained as the average of all the pairs 
directions with weights calculated as a function of the image sizes, image 
lengths and relative angle between the major axis of the two images. This 
multi-telescope stereoscopic analysis has been developed and validated for the 
4-IACT array VERITAS.

\subsection{Angular resolution}
\begin{figure}
  \includegraphics[width=13.5cm]{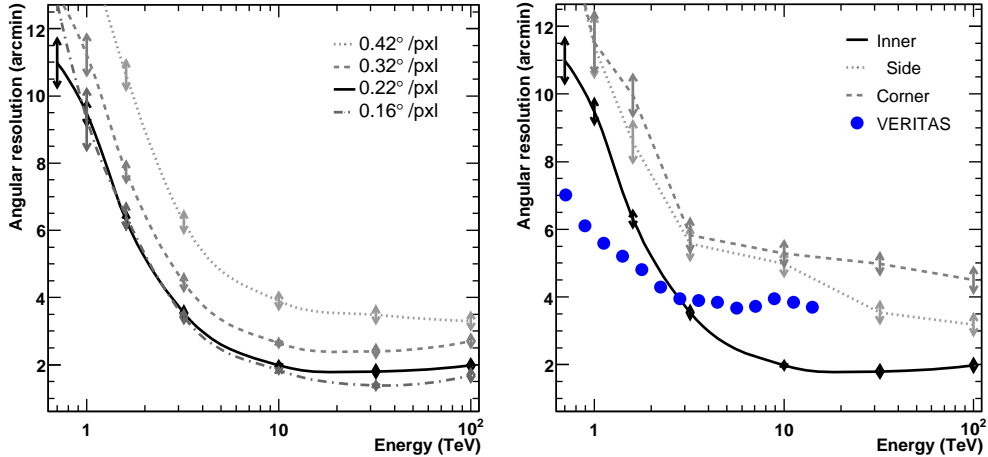}
\caption{Angular resolution, defined as the 68\%-containment radius, as 
a function of the $\gamma$-ray energy for (on the left) showers falling inside our
example array with different pixel size cameras and for (on the right) showers falling 
inside and outside of the array with $\omega=0.22^\circ$.}
\label{AngleRes}
\end{figure}

We first want to determine an optimal pixel size and continue our study 
with only one pixel size. In order to do this, we compared the angular 
resolution obtained for different camera pixelations. Figure \ref{AngleRes}
shows the $68\%$-containment angle of the point spread function as a function 
of the $\gamma$-ray energy for 30$^\circ$ zenith angle showers falling inside 
the array for different pixel sizes. There is a clear improvement between 
$\omega=0.42^\circ$ and $\omega=0.22^\circ$ but almost no improvement from
$\omega=0.22^\circ$ to $\omega=0.16^\circ$. As a consequence, we choose 
$\omega=0.22^\circ$ for the rest of this study. Smaller pixels would results in
rapidly increasing camera costs which are not justified by performance
improvements. It should be noted that even $\omega=0.32^\circ$ are resulting
in a very attractive angular resolution, better than 3$^{\prime}$ above 10\,TeV.

Figure~\ref{AngleRes} also shows the $68\%$-containment radius for 
the array inside and outside regions with $\omega=0.22^\circ$ and
compares them to the angular resolution of VERITAS~\cite{jamie08}.
The angular resolution of the inner region improves rapidely with the energy.
Above 10\,TeV, it is better than 2$^{\prime}$. Our estimation is
even conservative as only 12 telescopes are considered while more 
telescopes would in fact detect such high energy showers.
The angular resolution becomes better than what is achieved with current telescope arrays 
above 2-3\,TeV while our chosen pixel size is quite larger.
This is due to the fact that in small arrays, a large fraction of events
are showers falling outside where all the telescopes view the shower 
from a similar view point, making the stereoscopic reconstruction less 
effective. With a large array as well, the direction reconstruction of 
showers falling outside is clearly not as good as for showers falling inside.
Selection of showers falling inside or close to the array would allow to 
achieve a very good angular resolution, but would reduce the effective area.
An IACT array with very large inter-telescope distances, and correspondingly large 
field of views to reach for the highest energies, could not achieve a so
good angular resolution. For example, the array with $\Delta T = 500$\,m, 
$\psi = 10^\circ$ and the same pixel size studied in \cite{funk09}, would not 
achieve an angular resolution better than the current experiments even with 
a cut on the core impact distance.

\subsection{Threshold and collection area}
\begin{figure}
  \includegraphics[width=8cm]{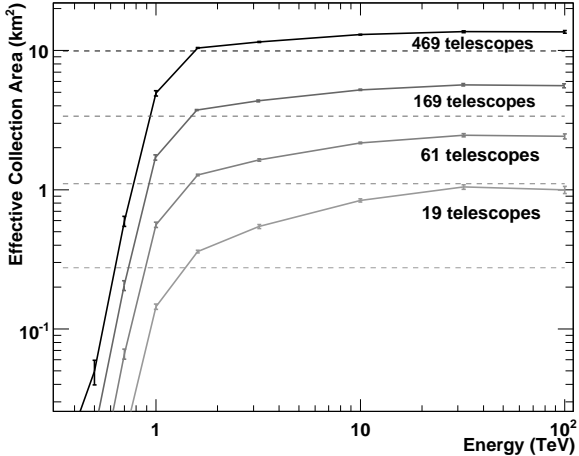}
  \caption{Effective collection area curves at 30$^\circ$ from zenith of the 
example arrays with different telescope numbers. Dashed lines show the geometrical 
area of the arrays (inner region).}
  \label{CollectionArea}
\end{figure}

We simulated $\gamma$-ray showers with energies from 350\,GeV to 100\,TeV at 
30$^\circ$ from zenith falling in a triangular cell and in the outside regions.
Half the 1\,TeV showers falling in a triangular cell are detected and pass 
our quality selection criteria. The energy threshold is around 
1\,TeV as we targeted when choosing the telescope diameter. 
For hexagonal arrays of various sizes, Figure~\ref{CollectionArea} shows the 
effective collection area $A_\gamma$ (at this level of event selection), as a 
function of energy. With a very large number of telescopes, 
$A_\gamma$ is dominated by the inside array region which provides
good angular resolution and result in a close to constant effective area 
above 1.6\,TeV. With a small number of telescopes however, $A_\gamma$ is 
dominated by the contribution from regions outside the array where the 
reconstruction performance  is not as good. As higher energy showers can be 
detected further away in the outside region, $A_\gamma$ depends on the
energy.  This effect is shower and telescope model dependent, thus opening up 
the possibility of systematic errors affecting the spectrum reconstruction.

As discussed earlier, 100\,TeV astronomy requires at least 100\,km$^2\cdot$hr
exposures. Generally IACT arrays are used to observe a given source not more
than 50\,hr per year. This sets a minimal requirement on the effective
collection area to $A_\gamma$\,$\simeq$\,2\,km$^2$ at 100\,TeV which is achieved 
by the 61-telescope array.
It should be noted that no more than 20 sources per year could be observed at the
required exposure level for 100\,TeV with the 61-telescope array while the present
TeV sky already counts more than 70 sources.
However, at lower energy, this array also offers a large collection area
which would satisfy the exposure requirement of Figure~\ref{exposure} in
a very short time. For example, in one hour, more than 25 events above 2\,TeV
from known sources would be collected. This would be enough for a detection
if it were not for CR-background-discrimination issues.

\subsection{Cosmic-ray discrimination and sensitivity}
The sensitivity of a ground-based $\gamma$-ray observatory is its capability 
to detect a $\gamma$-ray source hidden in an isotropic background generated by 
the CR air showers, and to measure the flux of this source. IACT can reject 
most of CR events thanks to an image shape analysis and, in the case of 
sources which are not too extended, a selection of events from the small sky 
region of interest.
The sensitivity depends on the observation time, the energy considered, the 
source characteristics (extension, spectrum) and the maximal flux-measurement 
uncertainty required. Here, we define the sensitivity as the minimum 
$\gamma$-ray flux $\Phi_\gamma(E_0)$ above a given energy $E_{0}$,
for a 5-standard-deviation detection based on more than 10 events in 50\,hr
of observation of a point-like source with a power-law spectrum with
an index $\Gamma=-2.5$.

In order to estimate the CR background, we need to determine the CR
acceptance $A_{CR}$ (collection area $\times$ solid angle) and the
CR-rejection efficiency. We simulated atmospheric Cherenkov signal
for proton and $\alpha$-particle showers from 500\,GeV to 500\,TeV, with
an energy distribution following a power law with spectral index of -2.7.
We estimate $A_{CR}$ with the same technique as for $A_\gamma$, subdividing
the array into triangular cells and outside regions. We selected the $\gamma$-ray
events by applying our standard analysis.
In principle, the $\gamma$-ray-like-event-selection cuts should be carefully 
optimized. This would require a huge amount of CR-shower simulations not 
available for this study. Instead of a careful optimization, we have chosen 
our selection criteria by eye, comparing image-parameter distributions for 
$\gamma$ rays and CR. The results presented 
here are therefore conservative. Using the CR flux measured with ATIC-2 \cite{atic-2}, 
we estimate the background rate of CR events passing selection cuts with a 
reconstructed energy above the given energy $E_0$ and a reconstructed
direction in the on-source region. The on-source region is defined as the 
68\%-containment radius for a $\gamma$-ray point source at energy $E_0$.
 
Generally, observations are taken in wobble mode with the studied source 
off-centered. The background level is then estimated from regions at the 
same distance from the camera center (off-regions). Here, we assume 
observation in wobble mode with an offset of 0.5$^\circ$ and a background 
determined from 7 off-regions. Then, we calculate the number 
of $\gamma$-ray events $N_{\gamma}$ needed for a 5 standard-deviation 
detection with the Li and Ma formula \cite{liandma}. We also required 
$N_{\gamma}$\,$\geq$\,10. We finally obtain the sensitivity $\Phi_\gamma(E_0)$
as following: \\

$\Phi_\gamma(E_0) = \frac{N_{\gamma} \cdot E_0^{\Gamma+1}} {\tau_{Obs} \cdot (-\Gamma -1) \times
 \int_{E_0}^\infty \epsilon_\gamma(E) \cdot A_\gamma(E) \cdot E^{\Gamma} \cdot dE}$ \\

where $\tau_{Obs}$\,=\,50\,hr, and $\epsilon_\gamma(E)$ is the $\gamma$-ray-selection
efficiency at the energy $E$. Figure~12 shows $\Phi_\gamma(E_0)$ for arrays of 61 telescopes
and 469 telescopes in comparison with the sensitivity of VERITAS \cite{jamie08}
and the Crab Nebula flux (considered here as a pure power-law spectrum with
a spectral index of -2.5).

\begin{figure}
  \includegraphics[width=8cm]{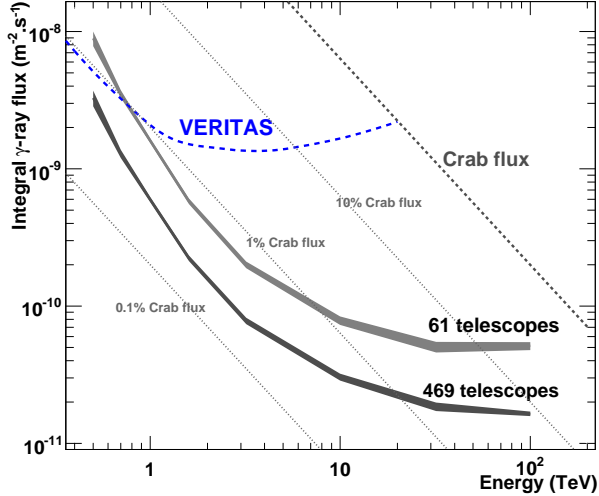}
  \caption{Sensitivity in 50\,h at 30$^\circ$ from zenith of the example array
  with different telescope number.}
  \label{sensitivity}
\end{figure}

Current IACT observatories achieve their highest differential spectrum sensitivity, 
$\sim$1\% Crab flux, between 300\,GeV and 1\,TeV. The 61-telescope example 
array could increase this energy range up to 10\,TeV and improve the best
sensitivity by a factor of two (0.5\% Crab flux in the 1.5-5\,TeV energy range).
Above 1\,TeV, the current IACT arrays get a good sensitivity because of a very efficient
background rejection but they are rapidly limited by the $\gamma$-ray statistics.
On the contrary, large IACT arrays obtain large statistics and their background
rejection improves rapidly with the energy. Sensitivity of our 61-telescope array 
is already $\sim$10 times better than VERITAS above 5\,TeV and $\sim$25 times above
10\,TeV. This sensitivity estimate is even very conservative because here the background
discrimination has not been optimized.
At higher energy, the sensitivity is much better than any existing experiment 
and one could measure the spectrum of a ($25\pm5$)\%-Crab-flux source above 100\,TeV
in 50\,hr. A dozen of known galactic sources should be measured up to 100\,TeV.
Many unknown sources with hard spectra may also be detected. For instance,
a 1\%-Crab-flux source above 250\,GeV with spectral index of -2 has a 20\%-Crab
flux above 100\,TeV.

With 469 telescopes, the sensitivity curve has essentially the same shape as
with 61 telescopes but with a factor of $\sim$\,3 improvement.
This means that with the 469-telescope array, one can reach about the same
sensitivity as with the 61-telescope array in one tenth of the time
(except the highest energies where the sensitivity is driven by 
the number of $\gamma$-ray events rather than by the CR-background rejection).
Then, one could study hundreds of sources per year in the 1-10\,TeV energy 
range.
Above 100\,TeV, the sensitivity is $\sim$\,8\%-Crab flux and, even if our study
for $\gamma$ rays stops at 100\,TeV, it is clear that such large
arrays would have useful sensitivities up to a few hundreds of TeV.

With larger telescopes or higher quantum efficiency cameras,
the sensitivity curve can be expected to essentially shift to the left. Energy 
threshold would be lower and the sensitivity would strongly improve below
10\,TeV. At the highest energy the array performances would remain more or 
less the same.

\section{Conclusion}
The next generation of IACT arrays is currently in design phase.
We investigated the performance of large uniform arrays of IACTs.
For the optimization,  we chose to maximize the effective collection
area for a given energy threshold and price. From considerations on the
atmospheric Cherenkov light properties and pricing, we found that the optimal
uniform IACT array is a hexagonal-lattice array with a 175\,m inter-telescope
spacing. The diameter of the telescopes depends on the targeted energy threshold
while the camera field of view depends on the maximal energy to be covered.
Additionally, provisions for the effective $\gamma$-ray field of view must also be
made. The $>$1\,TeV $\gamma$-ray astronomy is particularly interesting as it could
allow the identification of cosmic-ray accelerators up to the knee energy. Following
the optimal design prescriptions obtained, we designed a $\gamma$-ray observatory
dedicated to the 1\,TeV-100\,TeV energy range. We conservatively based our simulation
on existing telescopes, electronics and data analysis. We showed that an array of
61 telescopes of 3\,m diameter equipped of 5$^\circ$ field of view camera with 499
pixels would improve the sensitivity above a few TeV by an order of magnitude,
and could measure spectrum up to 100\,TeV for 25\%-Crab-flux sources
(a dozen of known galactic sources could reach this flux above 100\,TeV).
The large effective area achieved above a few TeV allows to reduce
the observation time without being limited by the $\gamma$-ray statistic
contrary to the current experiments at these high energies. This is important
as the observation time per source will decrease as the number of sources increases.
The very good angular resolution achieved at high energy ($\sim$\,2$^{\prime}$)
plays an important role for the morphology study and source disambiguation
as galactic sources are typically extended and concentrated in the galactic plane.

\section{Acknowledgment}
The authors are grateful to Pierre Sokolsky for his support and to Masahiro 
Teshima and the StarBase Utah team for informations about the 7 T.A. telescopes.

\end{document}